# Microfluidics Development at Berkeley

*Richard A. Mathies - University of California, Berkeley*

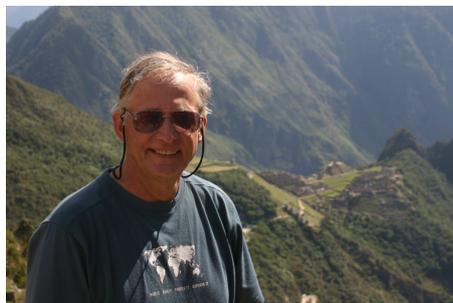

## Biography

*Richard A. Mathies is Professor Emeritus of Chemistry at the University of California, Berkeley. He received his B. S. in 1968 at the Univ. of Washington and Ph. D. in 1973 in Physical Chemistry at Cornell University. Following two years of postdoctoral study as a Helen Hay Whitney Postdoctoral Fellow at Yale, he moved to UC Berkeley in 1976. Mathies' biophysical research is focused on the use of resonance Raman and time resolved optical spectroscopy to elucidate the structure and reaction dynamics of energy and information transducing photoactive proteins. Mathies' work in analytical chemistry, biotechnology and the Human Genome Project led to the development of new high-speed, high-throughput DNA analysis technologies such as capillary array electrophoresis and energy transfer fluorescent dye labels for DNA sequencing. He also pioneered the development of microfabricated capillary array electrophoresis devices and microfabricated integrated sample preparation and detection technologies for lab-on-a-chip analysis systems applied to DNA sequencing, diagnostics, forensics, pathogen detection and space exploration.*

## Introduction

My first exposure to microfluidics occurred in the mid-1980's when we became aware of the microfabricated (by gelatin masked bead blasting) Joule-Thompson gas expansion cooling chips sold by MMR technologies for microscopes and detectors; a technology that came out of the lab of Prof. William A. Little at Stanford. These were arguably the first commercial microfabricated fluidic devices (1) that catalyzed my interest in the application of microfluidics to DNA analysis when the Human Genome Project began its technology development phase in 1987. This abstract will summarize briefly our journey through microfluidics over the past 30 years. My oral presentation will focus on only the most recent activities on technologies and devices for space exploration developed with the Berkeley Space Sciences Lab.

## Development of Microfluidics at Berkeley

I proposed using microfabrication for DNA sequencing to the DOE in about 1990, but my postdoc Xiaohua Huang (from Dick are lab) decided it would be faster to bundle conventional capillaries into a planar array thereby inventing Capillary Array Electrophoresis (2). He was right and this effort lead to the first commercial high throughput DNA sequencing instrument (Molecular Dynamics MegaBACE), energy transfer DNA sequencing labels, the "genome wars" between Molecular Dynamics/Amersham and Applied Biosystems, and the sequencing of the Human Genome in 2003.

Adam Woolley joined the group in 1992 and informed by the recent work of Harrison, Manz and Widmer embarked on the development of chips and methods to perform DNA analysis using microfabriated glass devices. He was rapidly successful publishing the first fragment sizing on chips (3), the first high-speed DNA sequencing on chips (4), and the first integration of a PCR reactor with a microfabricated separation system (5). It was evident that microfluidic technology offered a lot of potential for both sample processing and analysis but more robust microfabrication methods were needed to move from "one-off demonstrations" of functions to truly useful devices.

To this end Peter Simpson joined the group and focused on developing robust manufacturing methods providing devices with high quality channels and defect free surfaces along with micron precision to enable high-density devices (6 and Figure 1). This work ultimately led to the ability to fabricate and operate high-density 96 electrophoresis channel (6 inch) and remarkable 384 channel (8 inch) devices that remain to date a record for the field (7). All of these devices were detected using the high sensitivity confocal scanning systems pioneered by Jim Scherer in the lab.

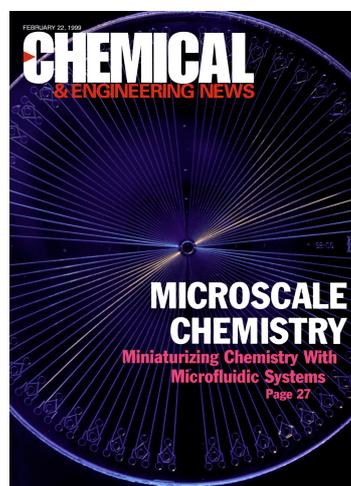

*Figure 1. Image of a 96-channel microfabricated capillary array electrophoresis device for DNA separations.*

Erik Lagally joined the group in 1999 intent on fabricating a monolithically integrated submicroliter volume PCR reactor with a microcapillary separation system. He succeeded, and his work was published as one of the first high impact papers in the then new journal *Lab Chip,* thereby helping it to be successful. A limitation of the Lagally work was the use of manually configured membrane valves for fluidic control. To address this limitation Will Grover began about the same time to develop a more robust microfabricated valve system that did not have the problems of typical MEMS valves. He developed a normally closed PDMS membrane valve structure that could be easily microfabricated (9). Grover and Jensen then realized that arrays of individually addressable valves (called automatons) could be used as programmable chemical processors that enabled valving, storage, mixing and pumping functions for all types of chemical analysis (10).

## Current activities

Current applications of our chip technologies for forensic identification (11) include the commercialization by IntegenX Inc. of microfabricated devices to perform real-time PCR amplification and analysis of STR's for Rapid-Hit forensic identification by military and civilian law enforcement. These devices are now starting to be used in police stations world wide for real time human identification thereby changing the paradigm for criminal identification.

Another exciting application is the use of microfabricated sample preparation, processing and analysis methods for the chemical exploration of our solar system (12). In particular, we are interested in looking for chemical signs of life such as chiral amino acids in the icy moons of Saturn and Jupiter where the need for autonomous processing and analysis in an extreme invironment is most important. For a more detailed review of these technologies (Figure 2) and how they can be integrated for space exploration see the recent review by Kim and Stockton (13).

The sky is not the limit for microfluidic systems.

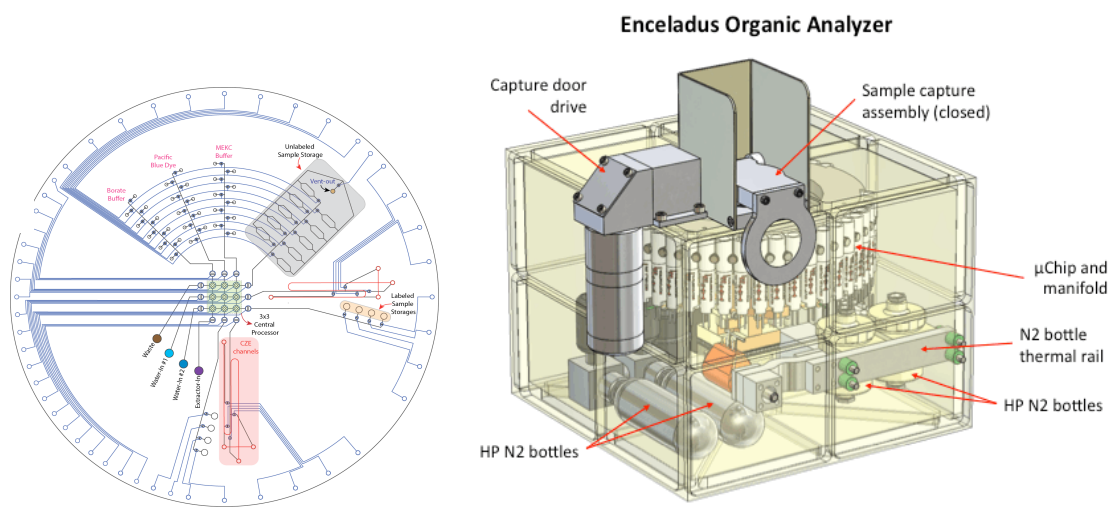

*Figure 2. Layout of a microfluidic chip for performing analysis of amines including amino acids and their chirality gathered from ice samples at the icy moons Enceladus and Europa. The central 3x3 automaton performs sample mixing and processing, there are two redundant folded electrophoresis channels for analysis, and reagents are stored dry in the hub and spoke channels (13). The image on the right presents the design developed by the Berkeley Space Sciences Lab of the Enceladus Organic Analyzer instrument containing this chip.*